\documentclass[]{spie}  

\usepackage{amsmath,amsfonts,amssymb}
\usepackage{graphicx}
\usepackage[colorlinks=true, allcolors=blue]{hyperref}

\title{Supporting users in their observation preparation - the ESO ObsPrep tool}

\author[a]{Monika G.\ Petr-Gotzens}
\author[a]{Vincenzo Forchi}
\author[b]{Andrea Mehner}
\author[a]{Thomas Bierwirth}
\author[a]{John Pritchard}
\author[a]{Ahmed M. Khan}
\author[a]{Olivier Hainaut}
\author[a]{Kevin Knecht}
\author[a]{Andre Graca}

\affil[a]{European Southern Observatory, Karl-Schwarzschild-Str.2, 85748 Garching, Germany}
\affil[b]{European Southern Observatory, Alonso de C\'ordoba 3107, Santiago 19, Chile}

\authorinfo{Further author information: (Send correspondence to M. Petr-Gotzens - E-mail: mpetr@eso.org)}

\begin{document} 
\maketitle

\begin{abstract}
Ground based observatories are experiencing significant technological advances and telescopes are getting equipped with highly complex instruments to enable cutting-edge astronomical research. The increasing complexity of such instruments displays a challenge to their users when they want to plan and prepare specific observation programmes, that often request to define large amount of detailed instrument-specific information. Therefore, at ESO we developed an Observation Preparation tool (in short ObsPrep tool), which provides a layer of abstraction over the instrument-specific technical details and tries to harmonize the users' experience when being faced with the preparation of observation blocks for different complex instruments. The ObsPrep tool allows interactive planning of the observations by visualization of various actions and observing strategies. In a joint approach by software engineers and astronomers, ObsPrep was started as a stand-alone java desktop application and over the years has evolved into an angular based web-application. It is now fully integrated in ESO's general use phase2 user interface (p2) that allows ESO users to create, modify, or delete observation blocks. 
Here we present the core functionalities of ObsPrep, as well as its special features and how the user interface integrates into ESO's generic phase2 observation preparation tool. We also outline the upcoming plans to incorporate the observation preparation for the instruments of the next generation of ESO's large telescopes, the ELT, into ObsPrep.
\end{abstract}


\section{INTRODUCTION}
\label{sec:intro}  

Contemporary astrophysical questions are significantly driving technological advances at ground based astronomical observatories. Telescopes are getting equipped with highly complex instruments in order to enable cutting-edge astronomical research. The increasing complexity of such instruments often displays a challenge to their users, the astronomical community, when they want to plan and prepare their observing programmes. At ESO's Paranal Observatory observing programmes are carried out via the execution of observation blocks (OBs). Hence, the creation of OBs is the essential work that astronomers have to do in order to translate their astrophysical observing request into a sequence of actions for the telescope and instrument. This task is accomplished with the help of ESO's phase2 proposal preparation software (\citenum{Beccari}), p2 in short, which offers a standardized, same look and feel, user interface for preparing OBs for all VLT and VLTI instruments. With p2 users define all instrument parameters, as for example the observation filter and exposure time, the width of the slit for spectroscopic observations, the size of the imaging field-of-view (FOV), and many more. However, some of the most simplest parameters, such as an instrument position angle on sky for instance, often have very instrument-specific definitions. For example, to place the imaging detector of the FORS instrument at PA\_on\_Sky = $35\deg$ users must specify $-35\deg$ in the OB while to achieve the same PA\_on\_Sky for the MUSE instrument a user must type in $+35\deg$. These instrument details are confusing and cumbersome for users to handle. 
Moreover, several instruments require to run instrument specific software\footnote{https://www.eso.org/sci/observing/phase2/sm\_overview.html} to produce, for example, fibre- or spectroscopic mask configuration set-up files for multi-object spectroscopy that are then attached to the respective OBs. 
Until recently such software were developed as highly instrument-specific, stand-alone tools that users would have to download and install in addition to the general observation preparation software, 
confronting users with many different single-installation tools, often requiring different platform pre-requisites. Therefore, a bit more than 10 years ago, ESO initiated the development of an Observation Preparation tool, now called ObsPrep, that should unify all the different instrument-specific preparation tools and would harmonize the users' experience when being faced with the preparation of observation blocks for different ESO instruments.  In this paper we describe the functionality of ObsPrep, its features, and how the ObsPrep tool has been designed in a joint approach by software engineers and astronomers towards a user-friendly, easy to use observation preparation tool for all ESO VLT+VLTI instruments.


\section{The ObsPrep tool}
\begin{figure} [ht]
   \begin{center}
   \includegraphics[height=8cm]{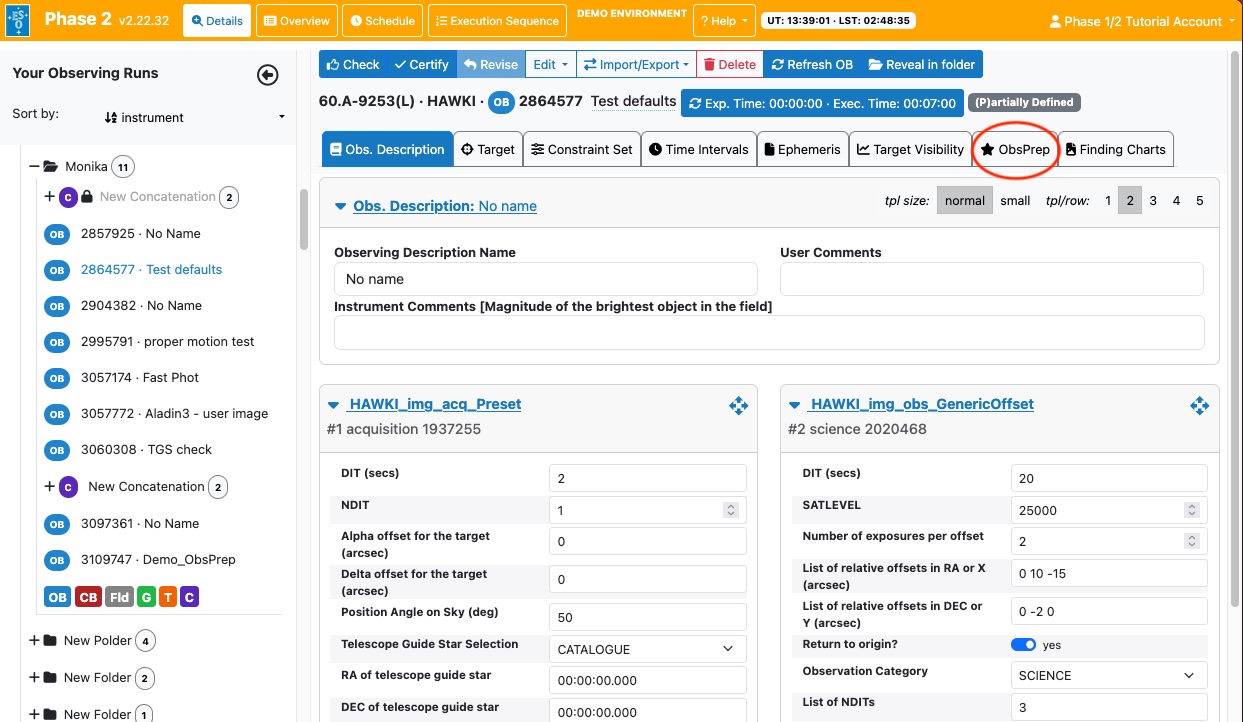}
   \end{center}
   \caption[ObsPrepP2] 
   { \label{ObsPrepP2} 
This figure shows the ObsPrep tool localization in ESO's observation preparation tool p2. The main window also shows the acquisition and observation template of a HAWK-I instrument OB.}
   \end{figure}

The ObsPrep tool started off as a standalone java desktop application which allowed interactive planning of the observations, e.g.\ users could fine-tune the science field pointing and were guided in the selection of blind offset stars, or AO guide stars, and/or the selection of VLT guide stars. By that time, ObsPrep could be used in tandem with ESO's general phase2 proposal preparation tool (p2pp), which as well was a java desktop client. When p2pp was migrated to a web-based application (p2web or simply p2) (\citenum{Bierwirth}) it naturally demanded to also migrate ObsPrep to the web and, even more important, to fully integrate it into p2. In Figure~\ref{ObsPrepP2} we show a snapshot of ESO's current p2 tool\footnote{https://www.eso.org/sci/observing/phase2/p2intro.html}, the official interface used by astronomers to prepare Phase 2 observing materials for VLT and VLTI instruments, with ObsPrep being accessible through one of the p2 tabs. p2 is used to create and manage Observation Blocks, design multi-OB observing strategies, make finding charts, and README files, and it directly links to the ESO operations database. An OB typically consists of one acquisition and one or more observing templates, and it is an indivisible unit of an observation defining all parameters necessary to point the telescope, set up the instrument and carry out the scientific observation without interruptions. 
   
\begin{figure} [ht]
   \begin{center}
   \includegraphics[height=8cm]{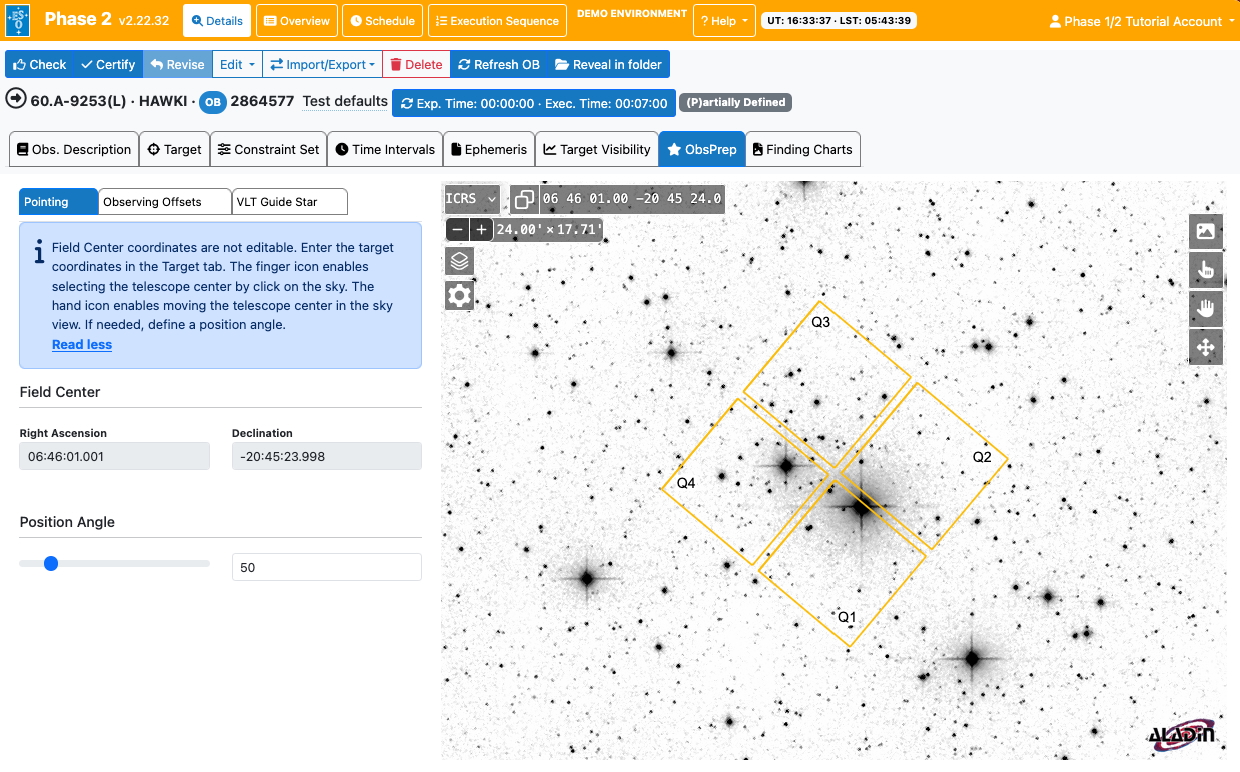}
   \end{center}
   \caption[ObsPrepP2] 
   { \label{HAWKIOB} 
The ObsPrep User Interface. Left: display of the science field center coordinates and the field position angle (PA), which can be adapted by using the slider or by direct input of the numerical value. Right: the ObsPrep SkyView showing the HAWK-I instrument detector footprint on a sky background image. Moving the PA slider will directly update the detector footprint in the SkyView as well as the respective PA keyword in the acquisition template.}
   \end{figure}

The scope of the ObsPrep tool is to guide the user in their choice of parameters and set-ups for the acquisition and observing template(s) of an OB. For this the general concept of ObsPrep is to 'visualize' the OB content and to allow interactive definition and selection of diverse OB parameters and parameter lists.  Figure~\ref{HAWKIOB} shows the 'landing' tab, that is the Pointing Tab, of the ObsPrep UI. Communication between p2 and ObsPrep is established by ObsPrep invoking the p2-api (application programming interface)\footnote{https://www.eso.org/sci/observing/phase2/p2intro/Phase2API.html} which keeps all information contained in p2 and the OB. In other words, when a user changes to the ObsPrep tab in p2, the ObsPrep tool 'grabs' basic information (via the p2-api) from the OB acquisition template, in order to decide which instrument and instrument-mode is defined, what to show in the graphical UI, and which interactive functionalities to offer accordingly. Having an OB with an acquisition template defined in p2 is hence the absolute minimum for ObsPrep to start working. In Figure~\ref{HAWKIOB} we show what is displayed by the ObsPrep UI in case of the simple HAWK-I OB seen in Figure~\ref{ObsPrepP2}. In the spirit of harmonizing the user experience an essential feature of ObsPrep is to map the many instrument specific keyword definitions of the templates into a generic definition for how to display and visualize setups in the ObsPrep UI, i.e. no matter which actual value for a field position angle is written in an instrument's acquisition template or what is a detector's actual default field orientation on the sky, the ObsPrep SkyView will always show a detector footprint such that North is up, East is left and with the astronomical definition of the position angle on sky, i.e.\ measured relative to the North.  

\begin{figure} [ht]
   \begin{center}
   \includegraphics[height=8cm]{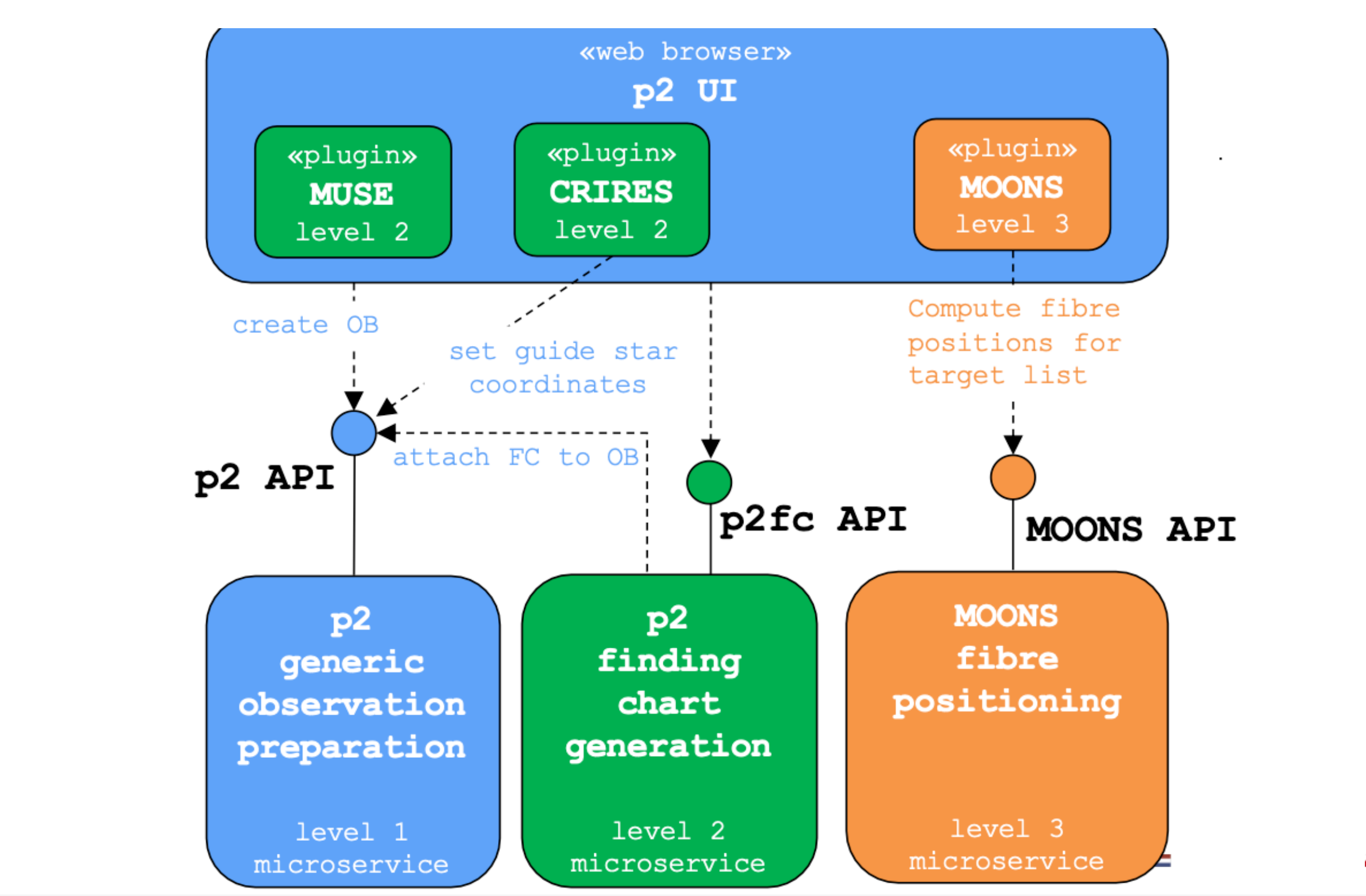}
   \end{center}
   \caption[example] 
   { \label{infrastrucure} 
The ObsPrep software infrastructure in the context of p2. Core functionalities are here referred to level~2 and marked in green, while level~3 are instrument-specific functionalities marked in orange.}
   \end{figure}

Overall, the ObsPrep tool is very versatile and allows interactive planning of the observations, e.g.\ the fine-tuning of the science field pointing, the selection of blind offsets or adaptive optics guide stars, the selection of VLT telescope guide stars, as well as visualization and planning of various observing strategies. The choices made are directly reflected by an update of the respective observing keywords and parameters in the associated OB. At this moment, ObsPrep is offered for all VLT instruments, although for KMOS only for standard star calibrations, and for FLAMES only for Argus fast acquisition. Support for VLTI instrument was added recently, starting with the GRAVITY “dual field wide”, and MATISSE and PIONIER coming soon. Technically speaking, ObsPrep integrates different back-end services, mainly featuring a common core functionality and a dedicated instrument-specific functionality for more complex observation preparation. A schematic overview of the software infrastructure is shown in Fig~\ref{infrastrucure}, while in the following sections we describe in detail ObsPrep's common core as well as its instrument-specific functionalities.

\subsection{Generic Core Functionality of ObsPrep}
The ObsPrep tool provides a set of generic core functionalities, that include all basic interactive planning capabilities, and that operate uniformly across all instruments. These functionalities are implemented through shared abstractions and interfaces rather than instrument‑specific logic. Although the underlying framework is common, certain features have instrument‑dependent configurations. For example, parameters such as the search radius or brightness limits for selecting candidate guide stars may differ between an infrared and an optical imager. These instrument‑dependent requirements are typically defined by the consortium or instrument operations teams and implemented in ObsPrep by ESO's software engineering department.

\begin{figure} [ht]
   \begin{center}
   \includegraphics[height=8cm]{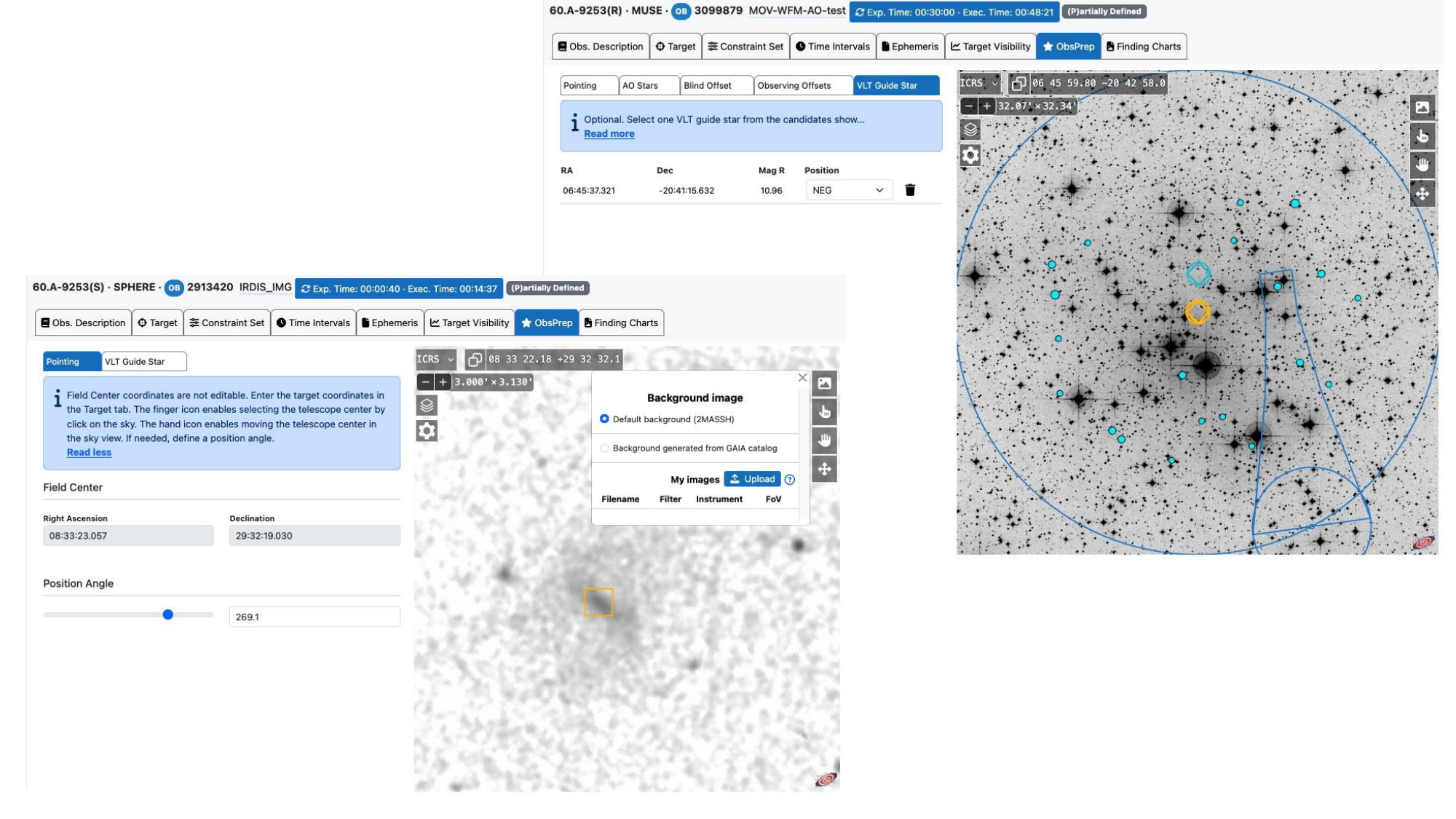}
   \end{center}
   \caption[example] 
   { \label{2inst} 
The ObsPrep/Pointing tab for a SPHERE instrument OB (left), and ObsPrep/VLT guide star tab for a MUSE instrument OB (right). Note the different number of ObsPrep tabs/functions for the SPHERE and MUSE OB.}
   \end{figure}

The ObsPrep tool is organized into multiple tabs, and the number of available tabs varies by instrument (see Fig~\ref{2inst}). However, the pointing tab is present for all VLT and VLTI instruments, while both the pointing tab and the VLT guide star selection tab are included for every VLT instrument. 

Within the {\bf pointing tab} (Fig~\ref{2inst}-left), the user is presented with the instrument’s field‑of‑view (FOV) footprint at the specified sky pointing position, which is defined by the OB target coordinates. The FOV geometry is overlaid on a background sky image. By default, this background is selected automatically based on the observing wavelength specified in the acquisition template. Users may, however, configure the background manually. Options include the default sky image, a synthetic background generated from the Gaia catalog (\citenum{gaia}), reduced data products retrieved from the ESO Science Archive (when available), or a user‑uploaded fits image. The pointing tab also enables adjustments to the FOV position angle and fine‑tuning of the science target pointing.

The {\bf VLT guide star tab} (Fig~\ref{2inst}-right) displays all stars suitable for use as telescope guide stars, together with their relevant properties. The geometric shadow of the guide probe arm is shown to indicate potential obscuration of the science field. When a user selects a candidate star from the sky view, the corresponding coordinates are transferred to the observation block, automatically updating the associated parameters in the acquisition template. 

\begin{figure} [ht]
   \begin{center}
   \includegraphics[height=5cm]{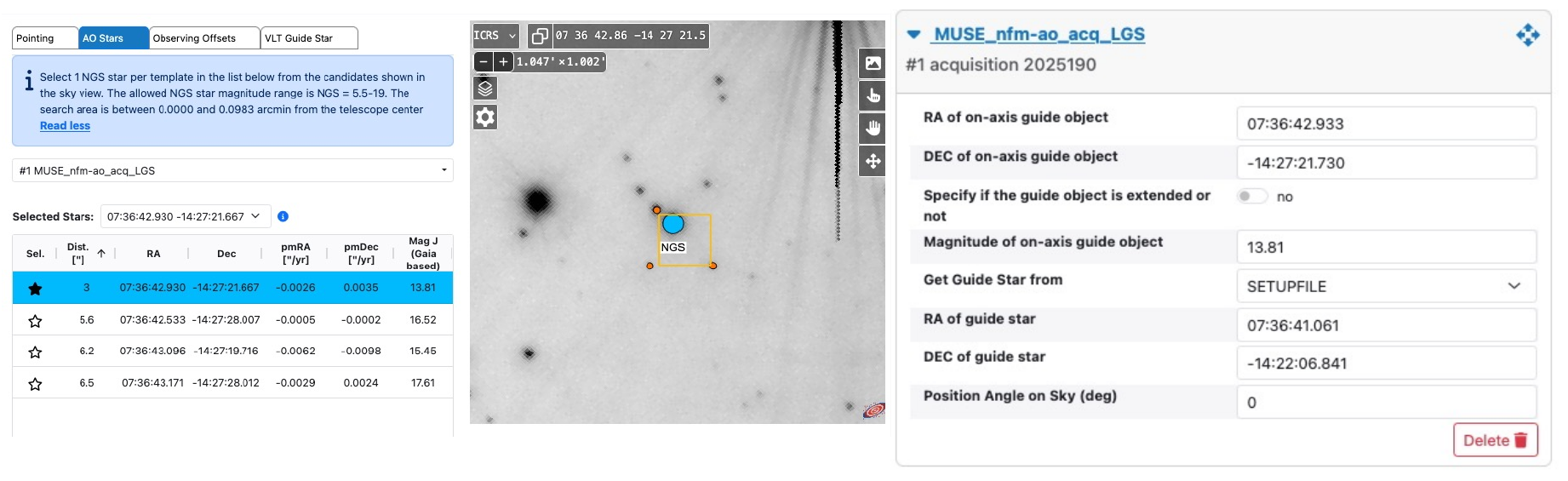}
   \end{center}
   \caption[example] 
   { \label{AOSTARS} 
Four AO star candidates presented by ObsPrep in a ranked table and indicated on the SkyView. The selected one is marked as {\it NGS }, and its coordinates are transferred to the OB acquisition template (right).}
   \end{figure}

Depending on the instrument and mode, the ObsPrep tool includes additional tabs that define core services. One of these is the {\bf AO Star tab}, which is available for instruments operating in adaptive‑optics–supported modes, such as MUSE, ERIS, or GRAVITY. This tab enables users to select suitable adaptive‑optics (AO) guide stars, including tip‑tilt stars or natural guide stars required by the instrument’s AO system. To identify potential candidates, ObsPrep issues a query to an astronomical catalogue—typically the Gaia catalogue, although the choice of catalogue is configurable—and returns all stars that meet the AO guide star criteria. These candidates are then displayed in the Skyview interface as coloured markers at their celestial coordinates, while a corresponding table lists them in order of increasing angular distance from the science target (Figure~\ref{AOSTARS}). The table also provides key properties such as stellar brightness, proper motion, and coordinates. Once the user selects one or more AO stars, their magnitudes, colours (when required), and coordinates are automatically transferred to the acquisition template of the OB.

The {\bf Blind Offset tab} is designed for cases in which the science target is too faint to be detected during acquisition. In such situations, the telescope is first pointed toward a bright, easily identifiable star, after which a precise offset is applied to reach the science field. The offset corresponds to the difference in right ascension and declination between the bright star and the target, expressed in arcseconds. While users would normally need to compute and enter these values manually, ObsPrep automates the process. By selecting a candidate offset star directly in the Skyview, the tool calculates the required offset and inserts the correct values into the appropriate acquisition‑template fields, significantly reducing the likelihood of user error.

Some instruments require additional specialised guide stars, addressed through the {\bf SlitViewer Guide Star (CRIRES) or Fringe Tracker Guide Star (VLTI‑Gravity+) tab}. ObsPrep queries the Gaia catalogue using predefined parameters and presents the resulting list of suitable guide stars. Users select a star interactively in the Skyview, and its relevant properties are automatically propagated to the OB acquisition template.

\begin{figure} [ht]
   \begin{center}
  \includegraphics[height=8cm]{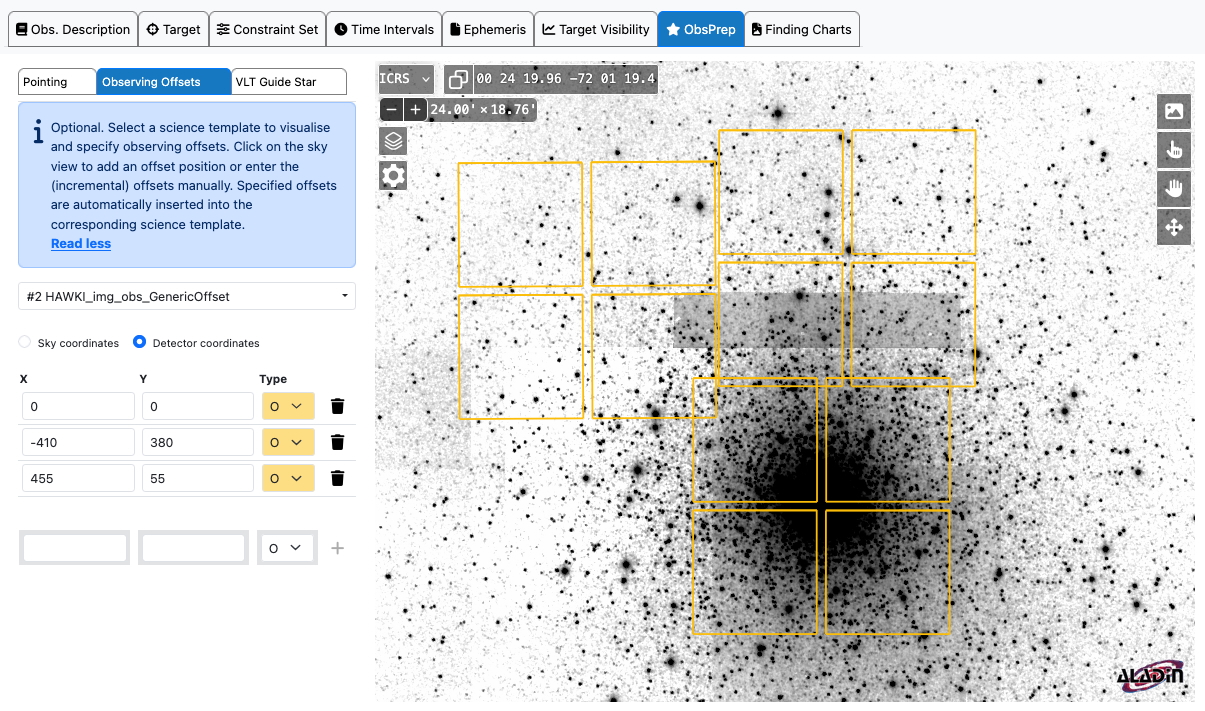}
   \end{center}
   \caption[example] 
   { \label{ObsOffsets} 
A pattern of 3 consecutive offsets defined in ObsPrep in order to image a large area of a globular cluster. }
   \end{figure}

The {\bf Observing Offsets tab} provides an especially powerful functionality within ObsPrep (Figure~\ref{ObsOffsets}). Observing offsets refer to changes in telescope pointing during an observation, enabling users to sample multiple positions within or around the science field. Through this tab, users can define offset sequences either by clicking directly on desired positions in the Skyview or by entering numerical offset values. ObsPrep immediately visualises the resulting pattern, allowing users to verify that mosaics, dithering patterns, or custom grids cover the intended sky region. This visual feedback is particularly valuable for complex observing strategies, such as constructing large mosaics, implementing sky‑background sampling patterns, or ensuring uniform spatial coverage across extended targets. Because offset conventions differ among VLT instruments—for example, a positive right‑ascension offset may correspond to an eastward move of the telescope for one instrument but a westward move for another—ObsPrep ensures that the correct sign and coordinate system are applied when populating the observing template. This automation greatly reduces the potential for configuration errors.

\begin{figure} [ht]
   \begin{center}
   \includegraphics[height=8cm]{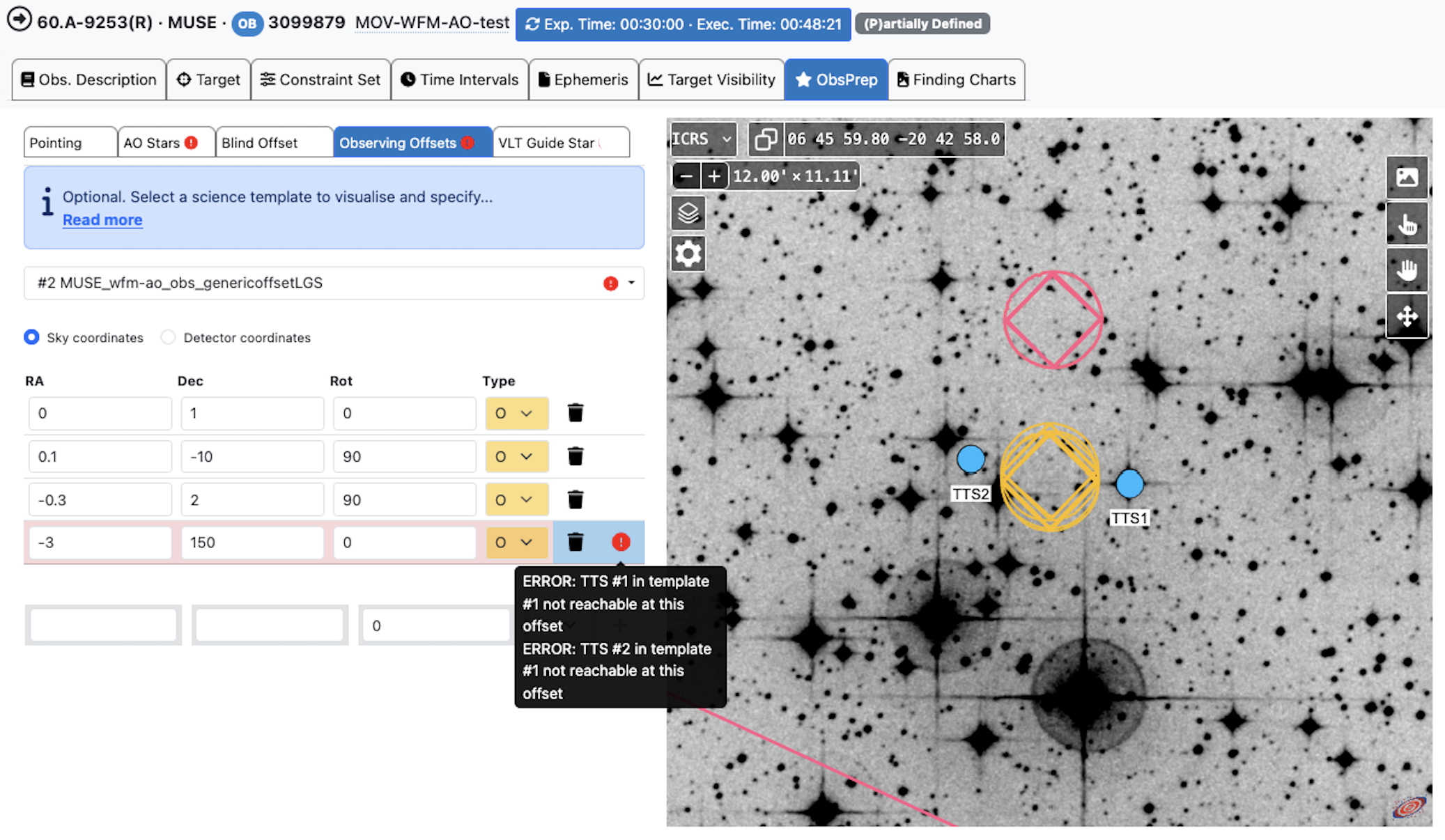}
   \end{center}
   \caption[example] 
   { \label{errorOffset} 
Feedback provided by ObsPrep when an error occurs. In this case, the fourth offset is too large in order to retain the TTS in the AO-patrol. }
   \end{figure}

ObsPrep also performs continuous consistency checks on the user’s configuration and provides immediate feedback on potential issues. This feedback appears as warnings and errors. Typical warnings include situations where no adaptive optics guide stars (or other guides stars) are found within the search region or cases where the catalogue query returns no entry at the position of the VLT guide star (or other guide star) coordinates as written in the acquisition template.
More critical problems generate error messages. These occur when required reference sources—such as the telescope guide star or AO star—become unreachable at one or more offset positions. This situation typically arises when the telescope offsets present in the Observing Offsets tab exceed the instrument’s patrol field, causing the guide star to fall outside the accessible region. Errors therefore highlight configurations that would fail during execution and require user correction (see Fig~\ref{errorOffset}). Users may also label offset positions as SKY or OBJECT. SKY positions, intended for background measurements, are exempt from reachability checks, allowing observers to construct offset patterns without being constrained by guide star availability at every position.
It is important to note that ObsPrep’s warnings and errors do not prevent submission of the OB. Users may still choose to ignore these messages and upload the OB to the ESO Phase 2 database. During submission, the OB is validated by the External Verification Module (EVM), whose checks partially overlap with—but are not identical to—those performed by ObsPrep. In this way, ObsPrep functions as an advanced diagnostic tool during preparation, while the EVM provides the authoritative verification required for final acceptance.

\subsection{Instrument-specific Functionality of ObsPrep}
\label{sec:ins-specific}
The instrument‑specific functionality of ObsPrep encompasses features that are unique to individual instruments and therefore cannot be accommodated within the common core services of the tool. These functionalities require dedicated processing logic and specialised algorithms, which are implemented as server‑side microservices. Each microservice can also be executed as a standalone command‑line application, ensuring flexibility for development, testing, and operational use. Responsibility for developing these microservices lies with the respective instrument consortium, which provides the software—typically written in C/C++, Java, or Python—for deployment and hosting at ESO in Garching.

Because these instrument‑specific components require tailored interaction modes, ObsPrep incorporates dedicated user interfaces (developed by ESO) for instrument-dedicated microservices, enabling users to run the instrument‑specific preparation software directly from within ObsPrep. All necessary input parameters are automatically extracted from the OB through the p2 API, complemented by data in configuration files and then passed to the microservice.

\begin{figure} [ht]
   \begin{center}
   \includegraphics[height=8cm]{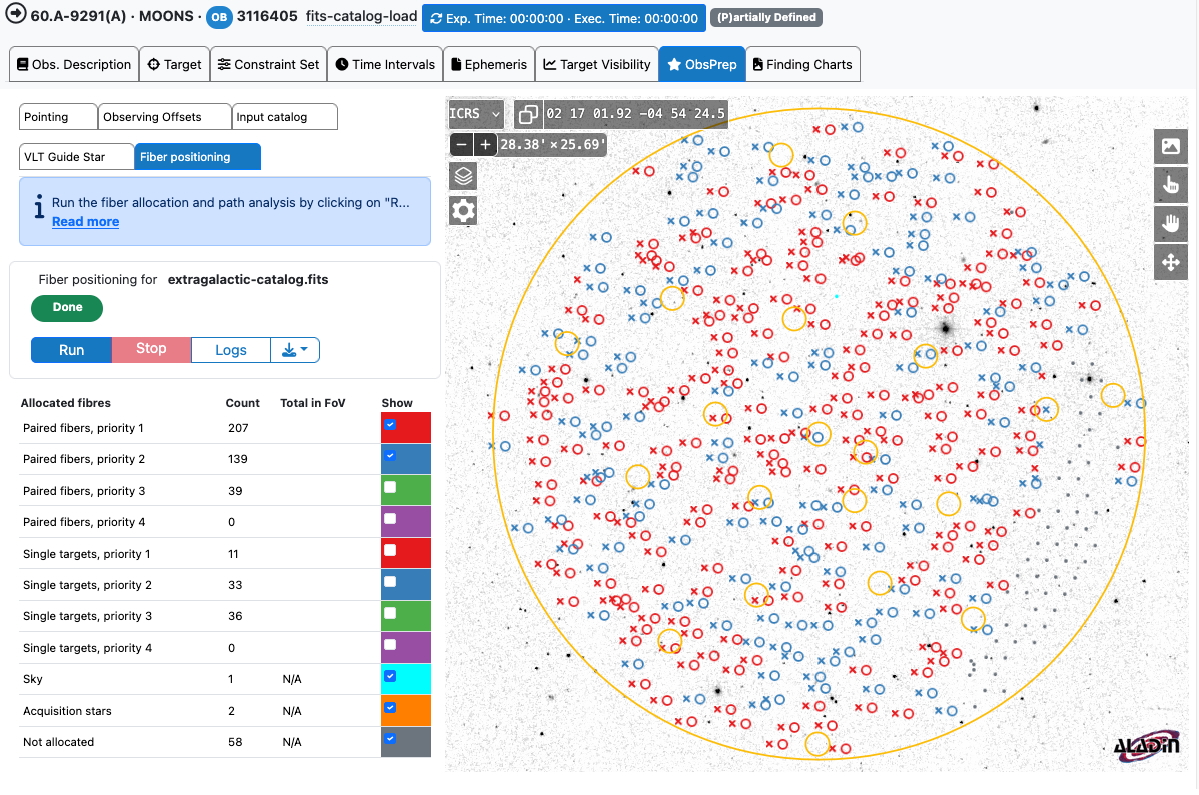}
   \end{center}
   \caption[example] 
   { \label{MOONS} 
Dedicated MOONS fibre positioning user interface, showing the configuration of paired fibres on SKy. The area where no fibres are seen, but targets marked with grey dots, is the obscuration by the VLT guide probe.}
   \end{figure}

A prominent example is provided by the MOONS instrument, which relies on a specialised microservice. This is the fibre‑allocation optimisation algorithm, known as MoonLight (\citenum{ML}), which maximises the number of fibres assigned to science targets, followed by the Path Analysis module, which computes the precise trajectory each positioner must follow to reach its allocated target. The ObsPrep UI for MOONS fibre positioning is shown in Figure~\ref{MOONS}. The output of the MoonLight microservice are a fibre‑configuration file and one or more motor‑path files. ObsPrep attaches these products automatically to the acquisition template of every MOONS observing block, ensuring that the instrument receives all required configuration data.

Another example is the TipTop microservice (\citenum{TipTop}), a fast Fourier‑based analytical simulation tool used to predict the point‑spread function (PSF) for instruments equipped with AO systems. TipTop computes the expected Strehl ratio at the science field centre or across the full two‑dimensional field, either for a single AO guide star or a full asterism selection. This capability provides observers with rapid, quantitative insight into the expected AO performance for their chosen configuration, and can be currently used in the ERIS instrument.

\section{Future Development of ObsPrep}

Future developments of ObsPrep will focus on expanding its capabilities to support observation preparation for all forthcoming VLT instruments, including MAVIS (\citenum{MAVIS}) and CUBES (\citenum{CUBES}). In parallel, work is underway to integrate the FORS Instrumental Mask Simulator\footnote{https://www.eso.org/sci/facilities/paranal/instruments/fors/doc/VLT-MAN-ESO-13100-2308\_P02.pdf} (FIMS) directly into ObsPrep. FIMS is currently a standalone tool that users must download and install locally, and it is required for generating configuration files that define target acquisition and focal-plane assembly setups for specific FORS observing modes. These include multi-object spectroscopy (both regular and polarimetric), spectroscopy with laser‑cut masks, and imaging observations employing occulting bars. Incorporating FIMS into ObsPrep will significantly facilitate its use and enhance the overall user experience. This integration is aligned with the commissioning schedule of the FORS-Up (FORS3) instrument, planned for 2027.

Observation preparation for ESO’s Extremely Large Telescope (ELT)—currently under construction—will require a fundamental revision. Each OB executed at the ELT must be fully specified and guaranteed to run without failure. Additionally, every exposure position must include a defined set of guide stars, and unlike VLT operations, the guide-star set may vary throughout the execution of a single OB. ObsPrep will therefore be responsible for selecting and validating the appropriate guide stars and/or guide star asterisms at each observing offset. Furthermore, unlike instruments at the VLT, each ELT instrument has its own requirements with respect to the telescope guide stars. While for the MICADO instrument (\citenum{MICADO}) there are no specific restrictions for the search area of ELT guide stars, for the METIS instruments (\citenum{METIS}) there will be "a blind zone" where the shadow casted by the ELT tertiary mirror blocks a portion of the sky, restricting the suitable guide stars area. Also, as the field rotates, any ELT guide star asterism will become unusable at certain LST ranges. All these interconnected dependencies must be considered for the implementation in ObsPrep.


Finally, the number and complexity of keywords in observing templates continue to increase, raising the risk of user confusion. Many templates now include intricate, instrument‑specific keywords. To address this, future versions of p2 will shift OB preparation toward an ObsPrep‑based abstraction layer that conceals most of the underlying template content within the p2 user interface. This will position ObsPrep as the central engine for constructing Observation Blocks, with the goal of simplifying the workflow for the users.

\bibliography{report} 

\begin{thebibliography}{1}

\bibitem{Beccari}
{Beccari}, G., {Bierwirth}, T., {Pruemm}, M., {Correia dos Santos}, P.~C.,
  {Forch{\i}}, V., {Brillant}, S., {Hainaut}, O.~R., {Marteau}, S., {Mehner},
  A., {Mieske}, S., {Mysore}, S., {Petr-Gotzens}, M.~G., {Pritchard}, J.,
  {Rejkuba}, M., {Tacconi-Garman}, L.~E., and {Wittkowski}, M., ``{Preparing
  observations for ESO telescopes: a versatile approach},'' in [{\em
  Observatory Operations: Strategies, Processes, and Systems
  IX}{\nolinebreak\hspace{0.1em}]},  {Adler}, D.~S., {Seaman}, R.~L., and
  {Benn}, C.~R., eds., {\em Society of Photo-Optical Instrumentation Engineers
  (SPIE) Conference Series} {\bf 12186},  121860N (Aug. 2022).

\bibitem{Bierwirth}
{Bierwirth}, T., {Amarandei}, B., {Beccari}, G., {Brillant}, S., {Dumitru}, B.,
  {Mieske}, S., {Pasquato}, M., {Pruemm}, M., {Rejkuba}, M., {Santos}, P.,
  {Tacconi-Garman}, L.~E., and {Vera}, I., ``{Flexible and dynamic observing at
  the ESO Very Large Telescope},'' in [{\em Observatory Operations: Strategies,
  Processes, and Systems VII}{\nolinebreak\hspace{0.1em}]},  {\em Society of
  Photo-Optical Instrumentation Engineers (SPIE) Conference Series} {\bf
  10704},  107041K (July 2018).

\bibitem{gaia}
{Gaia Collaboration}, {Vallenari}, A., {Brown}, A.~G.~A., {Prusti}, T., {de
  Bruijne}, J.~H.~J., {Arenou}, F., {Babusiaux}, C., {Biermann}, M., {Creevey},
  O.~L., and {Ducourant}, ., ``{Gaia Data Release 3. Summary of the content and
  survey properties},'' {\em Astronomy and Astrophysics}~{\bf 674},  A1 (June 2023).

\bibitem{ML}
{Franzetti}, P., {Belfiore}, A., {Gargiulo}, A., {Garilli}, B., {Beard}, S.,
  {Gonzalez}, O., and {Cirasuolo}, M., ``{SCROOGE, an Efficient Path-Planning
  Algorithm for the MOONS Spectrograph},'' in [{\em Astronomical Data Analysis
  Software and Systems XXXII}{\nolinebreak\hspace{0.1em}]},  {Gaudet}, S.,
  {Bohlender}, D., {Gwyn}, S., {Hincks}, A., and {Teuben}, P., eds., {\em
  Astronomical Society of the Pacific Conference Series} {\bf 538},  285 (Sept.
  2025).

\bibitem{TipTop}
{Neichel}, B., {Beltramo-Martin}, O., {Plantet}, C., {Rossi}, F., {Agapito},
  G., {Fusco}, T., {Carolo}, E., {Carl{\`a}}, G., {Cirasuolo}, M., and {Van Der
  Burg}, R., ``{TIPTOP: a new tool to efficiently predict your favorite AO
  PSF},'' in [{\em Adaptive Optics Systems VII}{\nolinebreak\hspace{0.1em}]},
  {Schreiber}, L., {Schmidt}, D., and {Vernet}, E., eds., {\em Society of
  Photo-Optical Instrumentation Engineers (SPIE) Conference Series} {\bf
  11448},  114482T (Dec. 2020).

\bibitem{MAVIS}
{Ellis}, S., {Zhelem}, R., {Horton}, A., {Chin}, T., {Fernando}, N., {Hewlett},
  M., {Lorente}, N., {Luo}, S., {McDermid}, R., {McGregor}, H., {Robertson},
  D., {Schwab}, C., {Smedley}, S., {Waller}, L., {Zheng}, J., {Kosmalski}, J.,
  {Cresci}, G., {Mendel}, T., {Bianco}, A., {Brodrick}, D., {Burgess}, J.,
  {Haynes}, D., {Greggio}, D., {Content}, R., and {Rigaut}, F., ``{MAVIS:
  imager and spectrograph},'' in [{\em Ground-based and Airborne
  Instrumentation for Astronomy X}{\nolinebreak\hspace{0.1em}]},  {Bryant},
  J.~J., {Motohara}, K., and {Vernet}, J. R.~D., eds., {\em Society of
  Photo-Optical Instrumentation Engineers (SPIE) Conference Series} {\bf
  13096},  130961B (July 2024).

\bibitem{CUBES}
{Cristiani}, S., {Alcal{\'a}}, J.~M., {Alencar}, S.~H.~P., {Avila}, G.,
  {Balashev}, S.~A., {Bastian}, N., {Barbuy}, B., {Battino}, U., {Calcines},
  A., {Calderone}, G., {Cambianica}, P., {Carini}, R., {Carter}, B., {Cassisi},
  S., {Castilho}, B.~V., {Cescutti}, G., {Christlieb}, N., {Cirami}, R.,
  {Coretti}, I., {Cooke}, R., {Covino}, S., {Cremonese}, G., {Cunha}, K.,
  {Cupani}, G., {da Silva}, A.~R., {De Caprio}, V., {De Cia}, A., {Dekker}, H.,
  {D'Elia}, V., {De Silva}, G., {D'Auria}, D., {D'Odorico}, V., {Diaz}, M., {Di
  Marcantonio}, P., {Fitzsimmons}, A., {Ernandes}, H., {Evans}, C.,
  {Franchini}, M., {Genoni}, M., {G{\"a}nsicke}, B., {Gneiding}, C.,
  {Giribaldi}, R.~E., {Grazian}, A., {Hansen}, C.~J., {La Forgia}, F.,
  {Landoni}, M., {Lazzarin}, M., {Lunney}, D., {Maciel}, W., {Marcolino}, W.,
  {Marconi}, M., {Migliorini}, A., {Miller}, C., {Noterdaeme}, P., {Opitom},
  C., {Pariani}, G., {Pilecki}, B., {Piranomonte}, S., {Quirrenbach}, A.,
  {Redaelli}, E.~M.~A., {Pereira}, C.~B., {Randich}, S., {Rossi}, S.,
  {Sanchez-Janssen}, R., {Seifert}, W., {Smiljanic}, R., {Snodgrass}, C.,
  {Squalli}, O., {Stilz}, I., {St{\"u}rmer}, J., {Trost}, A., {Vanzella}, E.,
  {Ventura}, P., {Verducci}, O., {Waring}, C., {Watson}, S., {Wells}, M.,
  {Wright}, D., {Zafar}, T., and {Zanutta}, A., ``{CUBES, the Cassegrain U-Band
  Efficient Spectrograph for the VLT},'' {\em The Messenger}~{\bf 188},  36--41
  (Sept. 2022).

\bibitem{MICADO}
{Davies}, R., {Schubert}, J., {Hartl}, M., {Alves}, J., {Cl{\'e}net}, Y.,
  {Lang-Bardl}, F., {Nicklas}, H., {Pott}, J.-U., {Ragazzoni}, R., {Tolstoy},
  E., {Agocs}, T., {Anwand-Heerwart}, H., {Barboza}, S., {Baudoz}, P.,
  {Bender}, R., {Bizenberger}, P., {Boccaletti}, A., {Boland}, W., {Bonifacio},
  P., {Briegel}, F., {Buey}, T., {Chapron}, F., {Cohen}, M., {Czoske}, O.,
  {Dreizler}, S., {Falomo}, R., {Feautrier}, P., {F{\"o}rster Schreiber}, N.,
  {Gendron}, E., {Genzel}, R., {Gl{\"u}ck}, M., {Gratadour}, D., {Greimel}, R.,
  {Grupp}, F., {H{\"a}user}, M., {Haug}, M., {Hennawi}, J., {Hess}, H.~J.,
  {H{\"o}rmann}, V., {Hofferbert}, R., {Hopp}, U., {Hubert}, Z., {Ives}, D.,
  {Kausch}, W., {Kerber}, F., {Kravcar}, H., {Kuijken}, K., {Lang-Bardl}, F.,
  {Leitzinger}, M., {Leschinski}, K., {Massari}, D., {Mei}, S., {Merlin}, F.,
  {Mohr}, L., {Monna}, A., {M{\"u}ller}, F., {Navarro}, R., {Plattner}, M.,
  {Przybilla}, N., {Ramlau}, R., {Ramsay}, S., {Ratzka}, T., {Rhode}, P.,
  {Richter}, J., {Rix}, H.-W., {Rodeghiero}, G., {Rohloff}, R.-R., {Rousset},
  G., {Ruddenklau}, R., {Schaffenroth}, V., {Schlichter}, J., {Sevin}, A.,
  {Stuik}, R., {Sturm}, E., {Thomas}, J., {Tromp}, N., {Turatto}, M.,
  {Verdoes-Kleijn}, G., {Vidal}, F., {Wagner}, R., {Wegner}, M., {Zeilinger},
  W., {Ziegler}, B., and {Zins}, G., ``{MICADO: first light imager for the
  E-ELT},'' in [{\em Ground-based and Airborne Instrumentation for Astronomy
  VI}{\nolinebreak\hspace{0.1em}]},  {Evans}, C.~J., {Simard}, L., and
  {Takami}, H., eds., {\em Society of Photo-Optical Instrumentation Engineers
  (SPIE) Conference Series} {\bf 9908},  99081Z (Aug. 2016).

\bibitem{METIS}
{Brandl}, B.~R., {Feldt}, M., {Glasse}, A., {Guedel}, M., {Heikamp}, S.,
  {Kenworthy}, M., {Lenzen}, R., {Meyer}, M.~R., {Molster}, F., {Paalvast}, S.,
  {Pantin}, E.~J., {Quanz}, S.~P., {Schmalzl}, E., {Stuik}, R., {Venema}, L.,
  and {Waelkens}, C., ``{METIS: the mid-infrared E-ELT imager and
  spectrograph},'' in [{\em Ground-based and Airborne Instrumentation for
  Astronomy V}{\nolinebreak\hspace{0.1em}]},  {Ramsay}, S.~K., {McLean}, I.~S.,
  and {Takami}, H., eds., {\em Society of Photo-Optical Instrumentation
  Engineers (SPIE) Conference Series} {\bf 9147},  914721 (Aug. 2014).

\end{thebibliography}
\bibliographystyle{spiebib} 

\end{document}